\documentstyle[12pt]{article}
\begin{document}
\baselineskip 10mm

PACS numbers: 36.40.Qv, 71.15.Nc, 71.15.Pd.

\vskip 4mm

\centerline{\large \bf Metastable quasi-one-dimensional ensembles of nitrogen
clusters N$_8$}

\vskip 2mm

\centerline{N. N. Degtyarenko, V. F. Elesin, L. A. Openov, A. I. Podlivaev}

\vskip 2mm

\centerline{\it Moscow Engineering Physics Institute
(State University),}
\centerline{\it Kashirskoe sh. 31, Moscow 115409, Russia}

\vskip 4mm

\begin{quotation}

By means of {\it ab initio} and tight-binding calculations it is shown that
metastable nitrogen clusters N$_8$ (boats) can form quasi-one-dimensional
ensembles in which the nearest clusters N$_8$ are bound to each other by
covalent bonds. Those ensembles are characterized by rather high energy
barriers ($\sim 0.3$ eV) that prevent the fission of the ensembles into
isolated N$_8$ clusters and/or N$_2$ molecules.

\end{quotation}

\vskip 6mm

A nitrogen molecule N$_2$ is known to be the most stable form of nitrogen at
standard conditions. Contrary to the overwhelming majority of other chemical
elements, the cohesive energy per atom, $E_{coh}(n)=E_{tot}(n)-E_1$ (where
$E_{tot}(n)$ is the total energy of an $n$-atom cluster N$_n$, $E_1$ is the
energy of an isolated nitrogen atom) decreases with $n$, so it is
energetically favorable for clusters N$_n$ with $n>2$ to decompose into
separated N$_2$ molecules and thus to lower their energy. However, recently
it was shown theoretically \cite{Ref1,Ref2,Ref3} that several nitrogen
clusters, e.g., N$_4$ and N$_8$, can exist in metastable states which
correspond to the local minima of the potential energy as a function of
atomic coordinates. Although metastable clusters N$_n$ with $n\geq 4$ have
not been synthesized yet, they attract much theoretical attention as
potential high energy density materials (HEDM). The energy
$E_{acc}(n) = E_{coh}(2) - E_{coh}(n)$ accumulated in the N$_n$ ($n\geq 4$)
systems approaches the value of $E_{acc}\approx 3$ eV/atom
\cite{Ref1,Ref2,Ref3}.

In order to make use of N$_n$ ($n\geq 4$) clusters as new efficient and
environmentally safe energy sources, one should be able to create bulk
metastable nitrogen structures composed of N$_n$ ($n\geq 4$) clusters. On the
one hand, in order such structures could exist, the nitrogen clusters should
be connected with each other by covalent bonds. On the other hand, those
bonds should not be "too strong" in order the clusters preserved their
individuality and the energy accumulated in such structures was not
substantially lower than the energy accumulated in an isolated cluster, i.e.,
the clusters should form ensembles. Besides, there should be a sufficiently
high energy barrier preventing the fission of ensembles into isolated
nitrogen clusters and/or nitrogen molecules.

An extensive theoretical study of Mailhiot et al. \cite{Mailhiot} suggests
that nitrogen can form bulk polymeric metastable structures at high
pressures. However, the polymeric structures cannot be considered as composed
of nitrogen clusters, i.e., as ensembles of clusters. In this paper, we
present the preliminary results of numerical simulations of bonding between
nitrogen clusters and show that boats N$_8$ can form metastable
quasi-one-dimensional ensembles in which the nearest clusters N$_8$ are bound
to each other by covalent bonds.

The version of the self-consistent-field theory with 3-21G, 6-31G$^*$, and
6-311G$^{**}$ basis sets in HyperChem was used in the present work. First, we
studied the geometry and energetics of isolated clusters N$_4$ (tetrahedron)
and N$_8$ (cubane and boat). Those clusters are shown in Figs. 1-3. The
accumulated energies are $E_{acc}(4)=(1.98\div 2.21)$ eV/atom for the cluster
N$_4$, $E_{acc}(8)=(2.44\div 2.59)$ eV/atom for the cubane N$_8$, and
$E_{acc}(8)=(1.53\div 1.66)$ eV/atom for the boat N$_8$, depending on the
basis set used. Our results agree with previous works \cite{Ref1,Ref2,Ref3}.

We have carefully analyzed a lot of different mutual arrangements of nitrogen
clusters. In both cases of N$_4$ and cubane N$_8$ clusters, we have found no
metastable configurations. However, in the case of boats N$_8$, we have
revealed that two boats can form a "cluster molecule" (N$_8$)$_2$. It is
shown in Fig. 4. The energy accumulated in this molecule, $E_{acc}(16)=1.79$
eV/atom (3-21G basis set), exceeds slightly the energy accumulated in an
isolated boat N$_8$. We have demonstrated by direct calculations that the
molecule (N$_8$)$_2$ is mechanically stable, i.e., all its oscillation
frequencies are real, giving evidence that the configuration shown in Fig. 4
corresponds to a local minimum in the total-energy surface as a function of
atomic coordinates.

In order to study the formation of large ensembles of boats N$_8$,
we employed a version of TBMD (tight binding molecular dynamics)
method with transferable interatomic potential \cite{Xu1,Xu2} used
by us earlier for carbon clusters
\cite{Openov,lowdim,Molecules,IWFAC,PLDS} and modified for
calculations on nitrogen systems. Although the TBMD method is not
as rigorous as ab initio approaches, it allows one to study
relatively large systems as well as to calculate directly the
lifetime of the metastable state and the height of the energy
barrier. For N$_4$ and N$_8$ clusters and (N$_8$)$_2$ "boat
dimers", the results of TBMD simulations agree qualitatively with
ab initio calculations. We have also shown that rather long
quasi-one-dimensional ensembles of boats can be formed through
attaching of new boats N$_8$ to the (N$_8$)$_2$ "molecule". The
ensembles (N$_8$)$_m$ with $m$ up to 6 have been shown to be
metastable with accumulated energy about 90 percents of the energy
$E_{acc}(8)$ accumulated in an isolated boat. The heights of the
energy barriers preventing the fission of metastable configurations
exceed 0.3 eV.

Finally, we note a recent success in recovering the non-molecular phase of
nitrogen at ambient pressure at temperatures below 100 K \cite{Eremets}.
Structure and physical properties of a new phase are not understood yet. Its
characterization remains an important issue. Presumably, it is polymeric.
This can be considered as a first step towards synthesis of bulk metastable
nitrogen structures, including "cluster phases" studied in this paper.

\vskip 6mm

{\bf Acknowledgments}

\vskip 2mm

The work was supported by the CRDF (project B-P/252),
the Contract DSWA01-98-C-0001,
and the Russian Federal Program "Integration" (Project AO133).


\newpage
\centerline{\bf Figure captions}
\vskip 2mm

Fig.1. Tetrahedron N$_4$.

Fig.2. Cubane N$_8$.

Fig.3. Boat N$_8$.

Fig.4. "Cluster molecule" (N$_8$)$_2$ composed of two boats N$_8$.

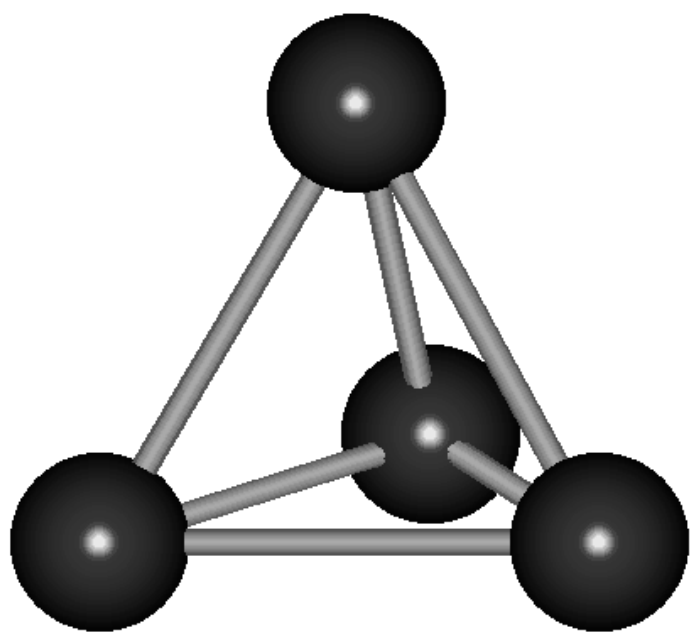
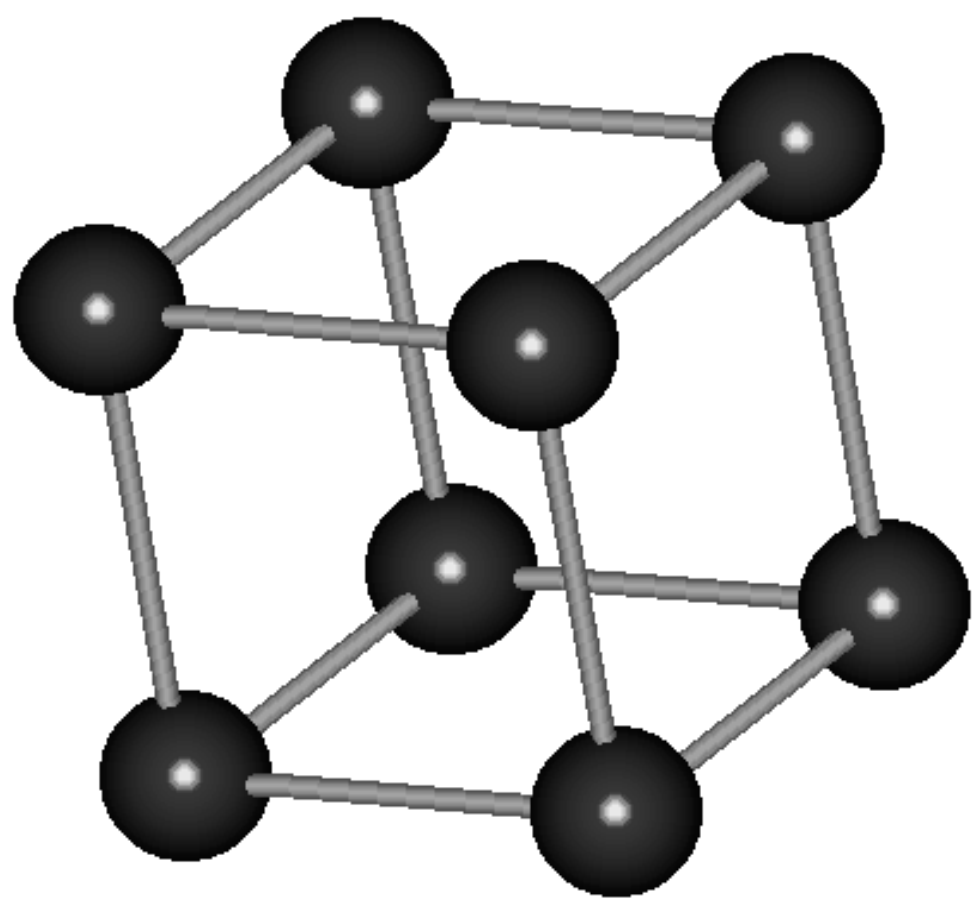
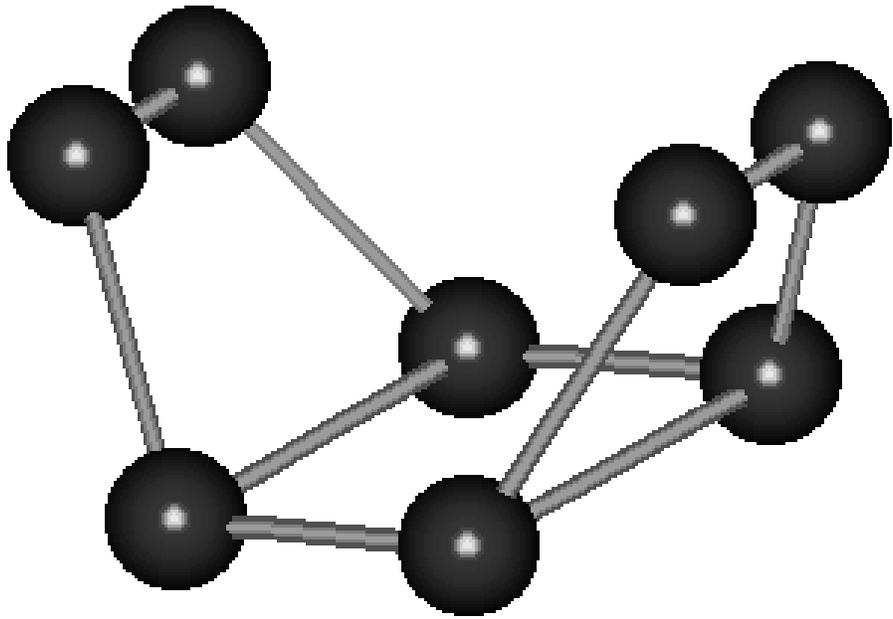
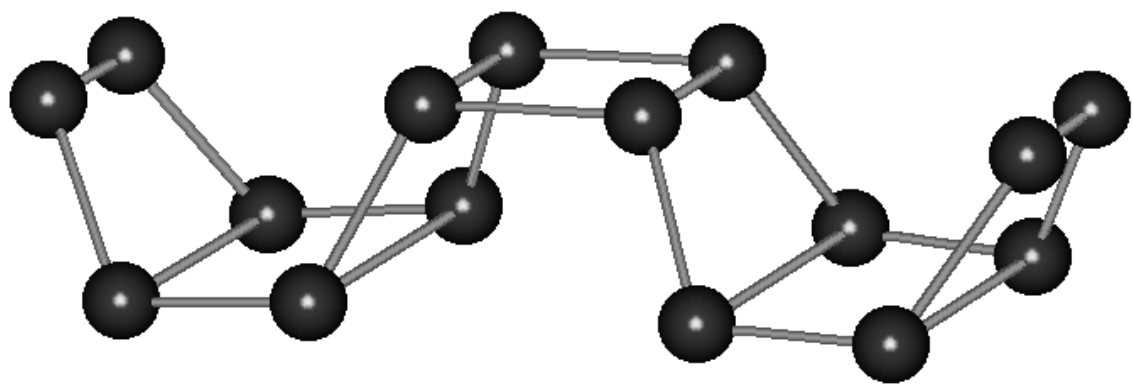
\end{document}